# How Air Entrapment in Hydrophobic Particle-Water-Air Mixtures Changes Post-Wildfire Mudflow Composition

Wenpei Ma[1] and Ingrid Tomac[2]
[1] Ph.D. Student, University of California San Diego, Department of Structural Engineering, La Jolla, CA
[2] Assistant Professor, University of California San Diego, Department of Structural Engineering, La Jolla, CA

**Abstract**
This paper shows critical new insights into how air entrapment affects the properties of rain-induced post-wildfire mudflows as a mixture of air bubbles, water, and hydrophobic sand. The idea of mudflows' internal structure containing trapped air bubbles is novel. Such mixtures can flow down slopes at incredible speeds, quickly blasting obstacles on the way and carrying large stone boulders and objects. The surficial soil particles turn hydrophobic due to the deposition of combusted organic matter during wildfires. Afterward, raindrops, splash, and erosion form devastating mudflows. We propose a new paradigm in which a significant amount of air remains entrapped in post-wildfire mudflow via hydrophobic particle-air attraction. Specific findings quantify the amount of air trapped within sand-water volumetric concentrations, the effect of intermixing energy, gravity, and sand particle size on outcome mudflow internal structure. As a result, little agglomerates of sand particles covering air bubbles characterize the mudflow mixture's internal structure.

**Introduction**
Post-wildfire mudflows are devastating natural disasters whose frequency increases with climate change and wildfire events. This paper shows how the mechanism of air entrapment into mudflow changes the properties and compositions of air-water-particle mixtures using laboratory experiments. First, wildfires in nature combust organic matter, litter, and other potential fuels present in soil and generate hydrophobic substances that precipitate and coat granular soil particles (DeBano, 1979; 1981; 1991; 2000; Neary et al., 2005). Then, rain erodes loose surficial layers that rill and blanket down slopes turning into catastrophic post-fire mudflows and can carry heavy gravels and boulders down the burned hillslopes and cause severe damage to lives and properties.

    Air entrapment into particle-water slurries is relevant for engineering applications as well as natural mudflows (Bull, 1963; Suhr et al., 1984; Römkens et al., 1997; Sheng et al., 2013; Tanaka et al., 2019; Cervantes-Álvarez et al., 2020; Dunkerley, 2021; Ong et al., 2021; Garoosi et al., 2022). Mudflow traps air from two sources: atmosphere air comes in when the mudflow stream moves down the tributary ravines and channels; otherwise, existing air from pores in the soil can be rolled up into the mudflow (Bull, 1963). Furthermore, debris flow can entrap air during impacting obstacles in the flow direction, affecting impact dynamics (Song et al., 2021; Garoosi et al., 2022). A numerical study reveals that the air entrainment rate in a granular flow is related to the gradient of solid velocity and flow thickness (Sheng et al., 2013). When a jet of grains falls into the water by gravitational forces, air entrapment is proportional to the volume of poured particles and inversely proportional to the granular size (Cervantes-Álvarez et al., 2020). In a similar experiment, particle hydrophobicity can enhance the ability to capture air bubbles so that fewer air bubbles flow up and out of the water body (Ong et al., 2021).

    Repulsive and attractive surface forces, collision mechanisms, and body forces govern the interaction between hydrophobic particles and air bubbles submerged in water. Surface forces are

repulsive van der Waals, repulsive electrostatic double-layer, attractive hydrophobic, and attractive capillary forces (Preuss and Butt, 1998; Gillies et al., 2005; Johnson et al., 2006). For example, the repulsive hydrodynamic force dominates the far-field of the hydrophobic particle-bubble interaction, while both repulsive and attractive surface forces govern near-field interactions (Ishida, 2007). However, the potential energy of attractive surface forces is one to two orders of magnitude larger than the repulsive double-layer van der Waals force (Lu, 1991). Furthermore, the attractive forces are proportional to hydrophobicity. Thus, the water film between an approaching particle and air bubble ruptures easily with hydrophobicity increase (Ishida, 2007). A liquid film between particle and bubble thins until critical and breaks during the hydrophobic particle-to-air-bubble attachment collision. Then, a three-point contact develops and progresses into a three-phase contact line (TPCL) (Nguyen et al., 1997). The bubble spreads over the particle surface until an equilibrium contact angle is reached (Fornasiero and Filippov, 2017). Hydrostatic pressure enhances the hydrophobic particle-bubble adhesion (Phan et al., 2003). For example, Fielden et al. (1996) show that hydrophobicity treatment of silica sand leads to a "jump" from repulsive to attractive interactions. The distance between the hydrophobic particle and bubble reduces compared to regular sand due to a hydrophobicity-promoted adhesion, leading to faster TPCL formation, consistent with capillary force mechanism at contact angles above 85°. However, despite hydrophobic surface alterations, repulsive forces remain present and the original electrical properties unchanged (Fielden et al., 1996).

Post-wildfire mudflow can move at a high velocity, up to 30 km/h (Cui et al., 2018), and this high speed may promote turbulence. The presence of particles in water generally can enhance or decrease turbulence (Gai et al., 2020). Turbulent flow regimes affect fluid drag, particle-bubble relative velocity, and particle dispersion rates, increase collision frequency and diminish the stability of hydrophobic particle-bubble attachment (Pyke et al., 2003, Liu and Schwarz, 2009). The submerged air-bubbles surface oscillates and loses sphericity at $Re>700$ (Schulze, 1989). As a result, an air bubble can rotate about the axis and induce a centrifugal force on particles or have an irregular trajectory or oscillate, while the particles fail to follow the bubble due to inertia in high-velocity fluid (Wang et al., 2016).

Gravity mainly contributes to particle-bubble detachment (Phan et al., 2003). For example, when a particle passes a bubble during falling (Verrelli et al., 2011) and slides towards the lower bubble hemisphere (Maxwell et al., 2012) or detaches (Gao et al., 2014). In some cases, gravity can promote attachment when a particle penetrates an air bubble at a certain grazing trajectory collision angle (Schulze, 1989). Furthermore, detachment of some hydrophobic particles from the air-bubble bottom surface occurs due to the viscous drag force from the counter-current fluid motion (Eskanlou et al., 2019) or lower localized particle surface-to-volume ratio at higher particle surface roughness and sizes (Fornasiero and Filippov, 2017). Additionally, for particles with a density similar to carrying fluid, the gravity is insignificant relative to inertia. However, if the particle density is larger than the fluid, the gravity effect surpasses the inertia effect (Brabcová et al., 2015).

Mudflow mixtures undergo complex interactions due to the attractive and repellent forces between three phases: water, hydrophobic particles, and air bubbles. Although many studies explain how a single or limited particle interacts with air bubbles, the macro-level effects and application of particle-bubble interactions with numerous particles and opportunities for bubble formation are not well understood. Previous research highlights the importance of solids availability on total air entrapment when grains enter the water, further enhanced by hydrophobicity. However, this study quantifies kinetic energy, phase ratios, mixing time, and hydrophobic particle size on entrapped air during fluid-particle-air mixing. Upscaling from micro to macro, understanding and quantifying the interplay between inertia forces, gravity forces, air-bubble formation dynamics, the collision between solid particles and bubbles, and solid particles' availability and physical properties lead to a better understanding of the amount of entrapped air and quantification post-wildfire mudflow mixtures density changes.

## Results
### Particle-Bubble Attaching and Detaching Dynamics during Mixing

Attachment, detachment, and collision processes like the previously identified literature occur during laboratory mixing within phase ratio and mixing kinetic energy ranges relevant for mudflows. The solid volumetric concentration can be as high as 60% in post-wildfire mudflows (Conedera et al., 2003; Cannon et al., 2001; Cannon et al., 2008; Kean et al., 2011; Cui et al., 2019; Lee & Widjaja, 2013). Submerged multi-particle and bubble mixtures undergo three main sub-processes during mixing: particle-bubble attachment, the collision between bubbles covered with particles, and particle-bubble detachment (Figs. 1a-c). Mixing enhances the interaction of initially separated solid particles, air bubbles, and water. Solid particles approach and attach to air-bubbles forming agglomerates. As the mixing process continues, agglomerates collide, some merge into larger agglomerates, and others vanish. At the same time, particles can also detach from the agglomerates. For example, Fig. 1a shows that only a few particles are attached to bubbles within the first 5 s of mixing, and the rest are still floating in the carrying fluid. At 20 s, more sand particles stick to air bubbles (Fig. 1b). Finally, particle-covered bubbles collide and form extremely large agglomerates. A wide initial distribution of across-size agglomerates reaches a stabilized size range equilibrium after about 50 s of mixing (Fig. 1c). Subsequent mixing beyond this point does not produce additional agglomerates for fine sands. Wang et al. (2016) observed three particle-bubble detachment mechanisms: centrifugal force on particles due to the rotation of the bubble about its axis in a vortex, irregular trajectories of the particle-bubble complex under motion, strong oscillation of the bubble surface which expels the particles. Experiments in this study demonstrate complex and varying particle-bubble trajectories. Besides, bubble surfaces sometimes exhibit a violent oscillation motion.

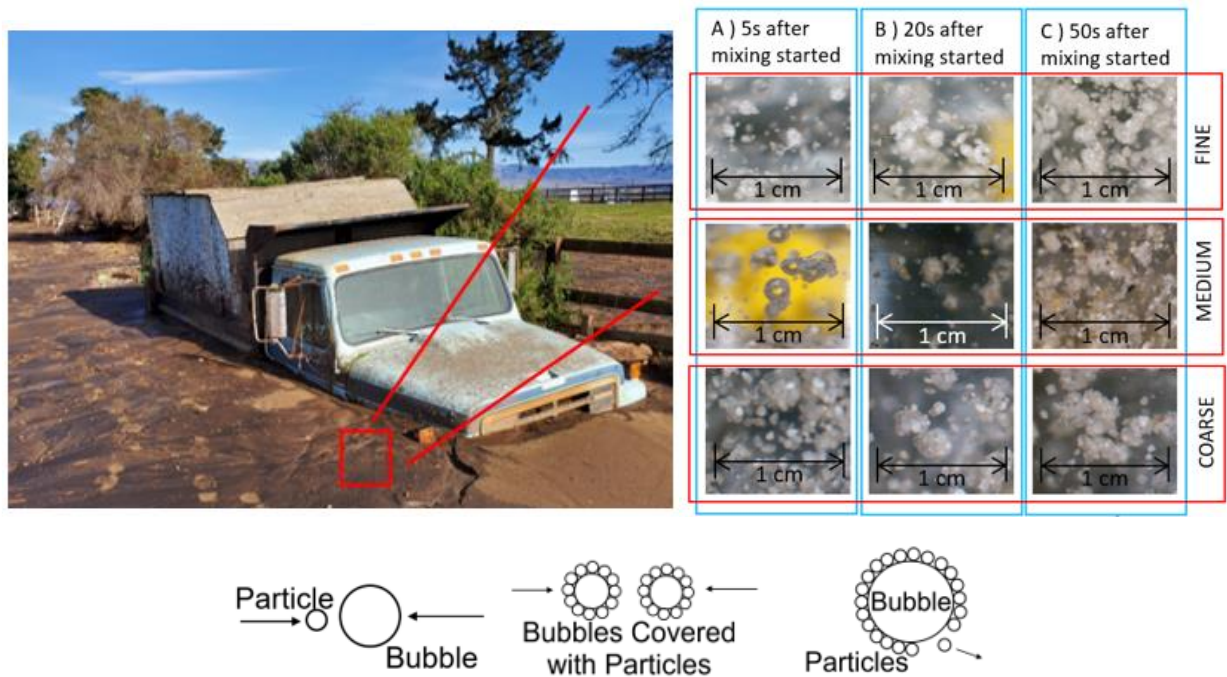

Figure 1. Different sub-processes of particle-bubble interaction include particle bubble approaching, attachment, interactions, and detachment. In experiments at 7.78 m/s mixing speed and initial solid volumetric concentration of 5%: A) within the first 5 s showing a diagram of particle-bubble attachment mechanisms, B) within 20 s showing a diagram of agglomerate-agglomerate collision mechanism, and C) within 50 s after mixing started, showing agglomerate-agglomerate collision and particle-bubble separation mechanism.

**Agglomerate Formation and Sizes, Overall Air Entrapment**

Among different factors like mixing time, mixing speed, initial solid volumetric concentration, and sand types, experiments show that the sand type factor strongly correlates with the final size of agglomerates. Therefore, coarser hydrophobic sands lead to larger agglomerates (Fig. 2). Furthermore, absolute agglomerate survival is a dynamic combating procedure between positive and negative forces. Fig. 3 shows the modified Bond Number relationship with the mean particle diameter and mixing speed. Coarser sand particles and higher mixing speed lead to a more significant modified Bond number, where agglomerates become more unstable, and particles detach from bubbles more easily.

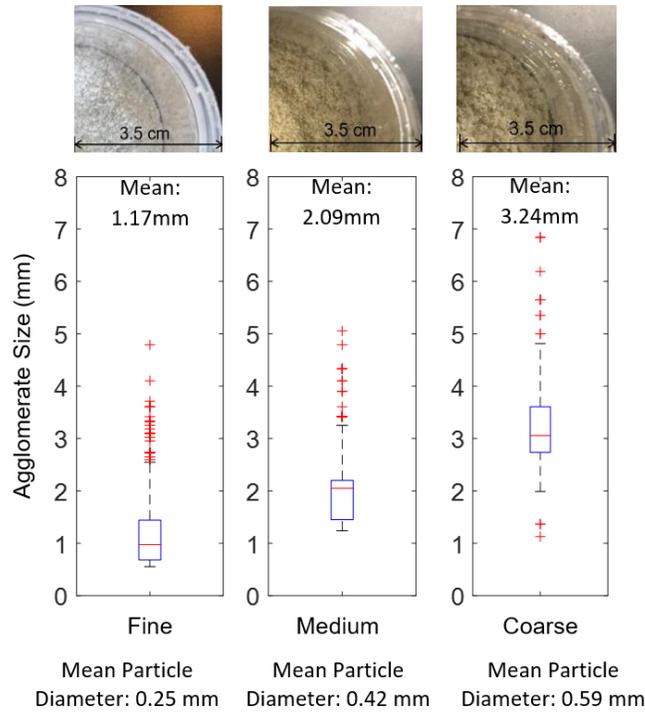

Figure 2. Effect of the sand particle size on the agglomerate size.

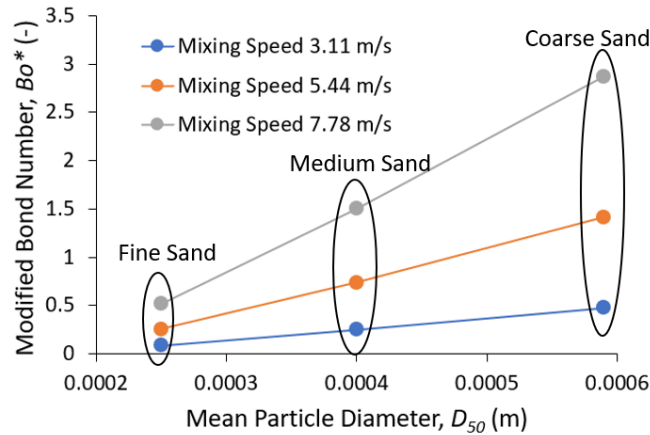

Figure 3. The combined effect of mean particle diameter and mixing speed on the modified Bond Number.

At a macro-scale, relative to the air entrapment can be expressed as the percentage of air or the ratio of air and the total volume of the mixture containing air, water, and particles ($V_a/V_{tot}$), shown in Figs. 4 and 5. However, the air-trapping overall is significant and more extensive than 40% relative to the volume of solids, which can be further explained by comparing the differences in the apparent void ratios, $e^*=V_a/V_s$. Figs. 4a-c shows the ability of the low, 5%-25% sand-in-water volume mixture to trap air under different mixing regimes. Relatively lower final air content is directly related to the amount of the hydrophobic particles, which is relatively low in a total mixture.

Experiments confirm (Figs. 4a-c) findings from Cervantes-Álvarez et al. (2020), who showed that an initial solid volumetric concentration plays an overwhelmingly dominating role in the final air trapping volume. Looking orthogonally to the $V_s/V_w$ axis in Fig. 4, which represents the initial solid volumetric concentration, one can always observe a significant ability to trap air phase, no matter what the other specific coupled condition is. Alternatively, when looking parallel to the $V_s/V_w$ axis, the effects of related factors have different significance levels on the ability to trap air phase. For example, trapped air steeply decreases with the sand particle size increase in Fig. 4a, a little bit less steep with the mixing speed increase in Fig. 4b, and gradually decreases and approaches an equilibrium as the mixing time progresses (Fig. 4c).

Additionally, the higher initial solid volumetric concentration helps to magnify the impact of the mean particle diameter, mixing speed, or time. In Fig. 4c, for example, while at the lowest initial solid volumetric concentration, the apparent void ratio increases from 11% to 48%. In addition, the apparent void ratio rises from 15% to 69% at the highest initial solid volumetric concentration.

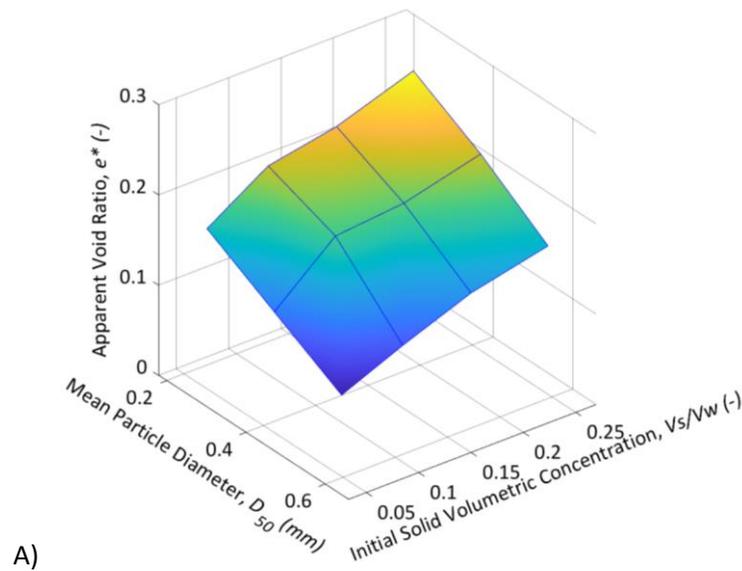

A)

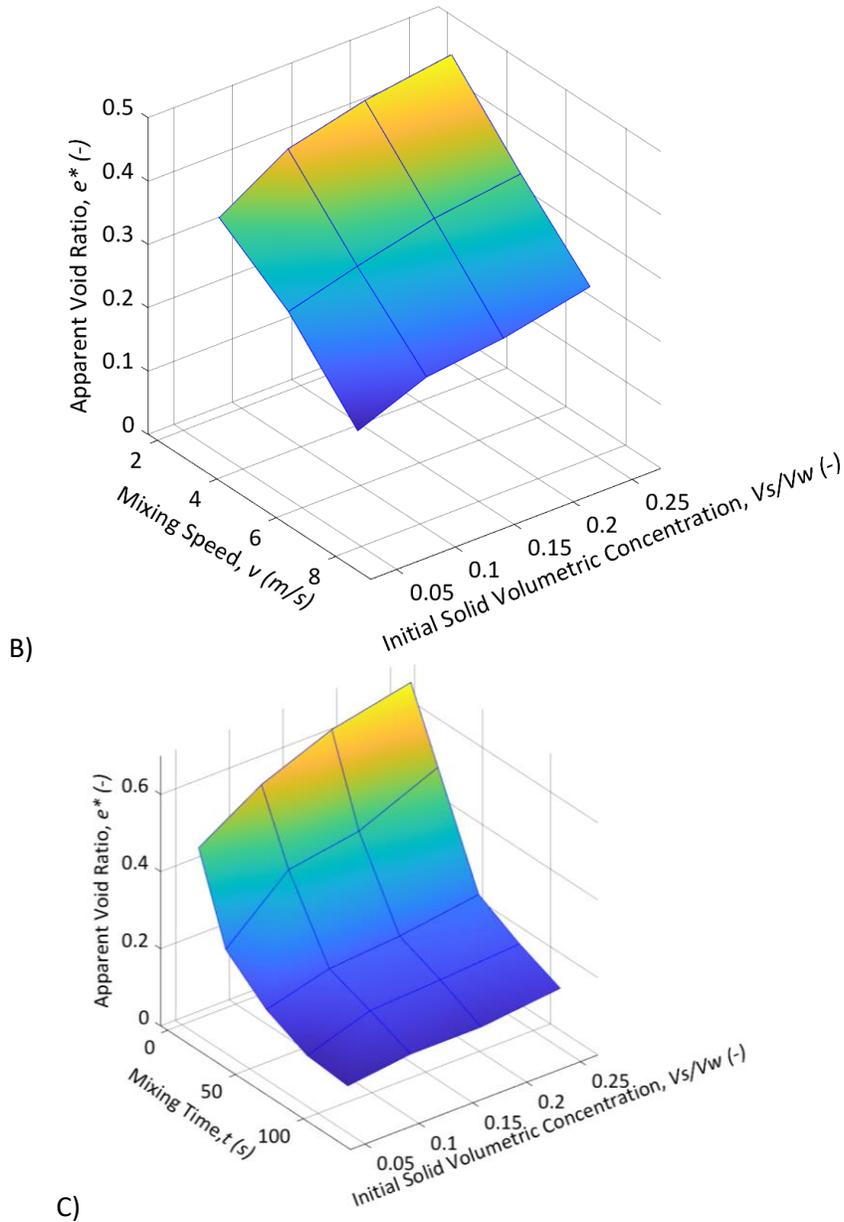

B)

C)

Figure 4. Coupled effect of initial solid volumetric concentration and other factors including A) type of sand particles, B) mixing speed, and C) mixing time on the ability to trap the amount of air phase. Results are shown in the intersection of mesh lines.

Since the effects of mixing time correspond directly to mudflow dynamics, it is interesting to perform an in-depth analysis of experimental results. Figs. 5a-d quantify how the prolonged mixing time decreases trapped air aiming towards equilibrium by expelling some amount of entrapped air bubbles. Figs. 5a-c shows that less observable agglomerates form when the mixture undergoes a long mixing. Mixing time and sand type have a coupled effect, and coarser sand needs less time to reach equilibrium. The mixing time effect couples with initial solid volumetric concentration, where the higher concentration causes a longer time to get an equilibrium state. Fig. 5d synthesizes coupled effects and serves as a base for the proposed empirical relationship:

$$\frac{e^*}{Vs/Vw} = 8.34(tD_{50})^{-0.59} \qquad (1)$$

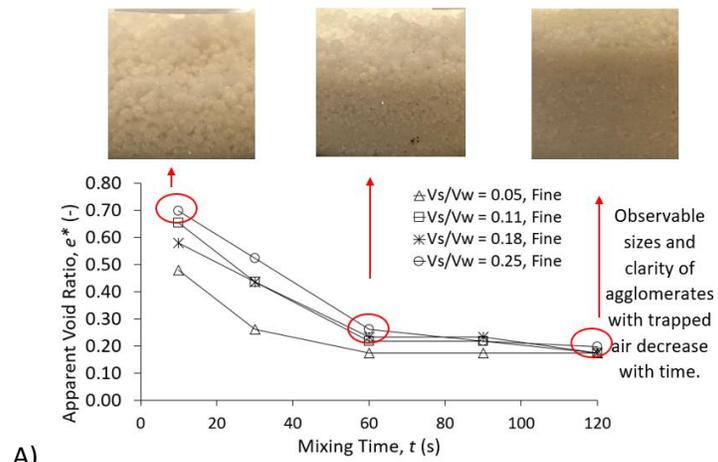

A)

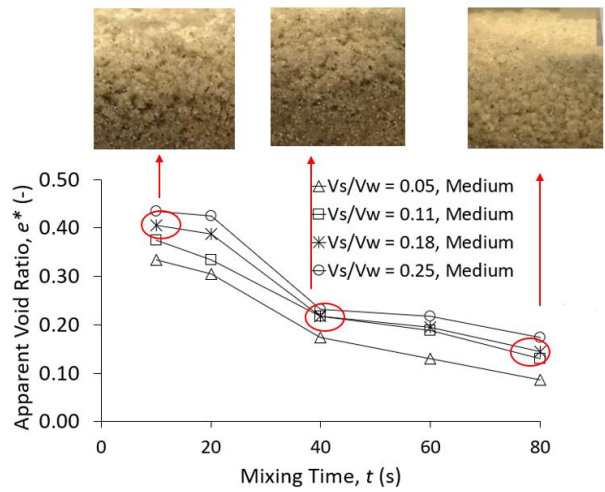

B)

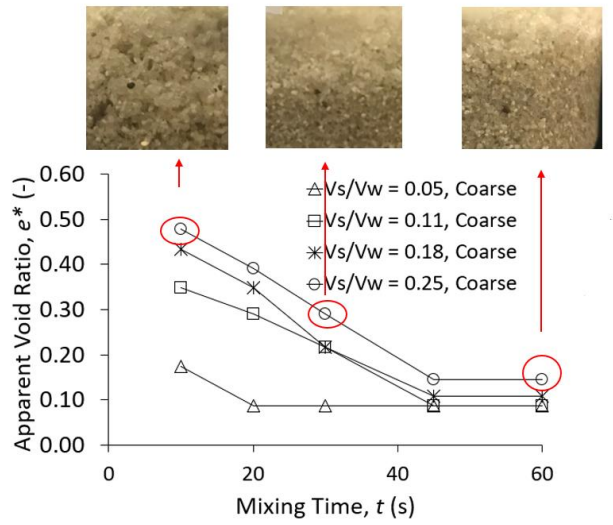

C)

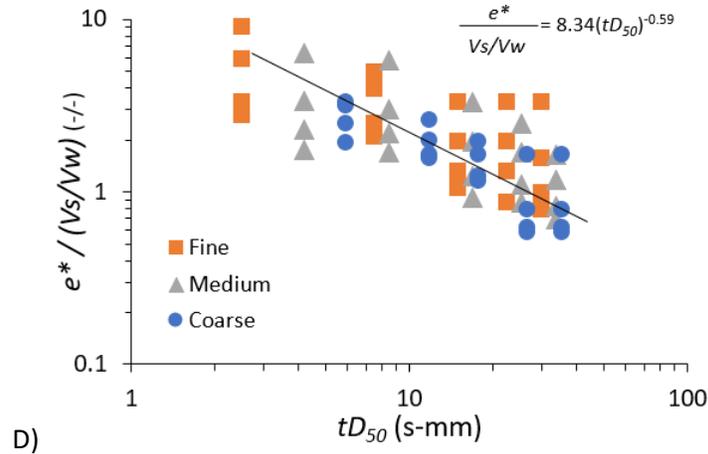

D)

Figure 5. Coupled effect mixing time and types of sand particles on the ability to trap air phase for A) fine sand particles, B) medium sand particles, D) coarse sand particles, and D) combined results.

High mixing speed leads to less air entrapment into the final mixture for all sands (Fig. 6). Results of the highest mixing rate (as circular data) show that the apparent void ratios are around 10% to 20%. In contrast, the results with the lowest mixing speed (demonstrated as triangular data) show that the apparent void ratios are around 30% to 40%. Besides, a higher amount of initially available solid will better capture air bubbles since all the lines in the figure have an increasing trend. At last, coarser sand has less ability to trap air bubbles while coupling with other factors. For example, for experiments at the lowest mixing velocity and all initially available amounts of solids, coarser sand (demonstrated in brown colors) has the smallest apparent void ratio. Fig. 7 shows another way of examining the effect of mixing speed. The figure describes an empirical correlation between air entrapment normalized by initial solid volumetric concentration and $Bo^*$. As demonstrated in Fig. 3, $Bo^*$ depends on the mean particle diameter. Therefore, we propose an empirical correlation:

$$\frac{e^*}{V_s/V_w} = -0.81 \ln B_o^* + 2.07 \qquad (2)$$

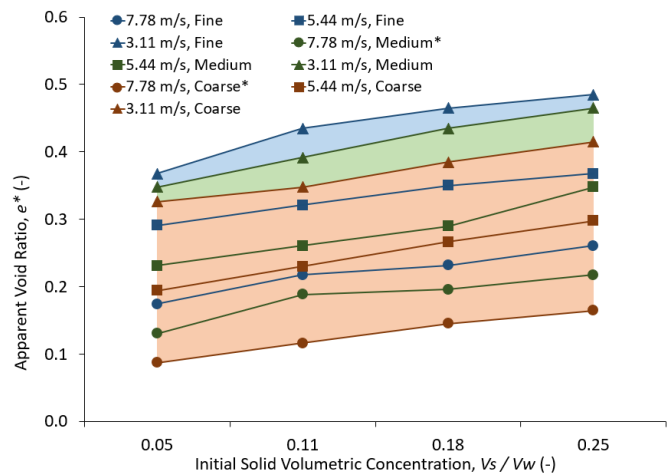

Figure 6. Coupled effect mixing speed and types of sand particles on the ability to trap air phase for different sand particle sizes. The Star sign indicates the average results for multiple tests under the same conditions.

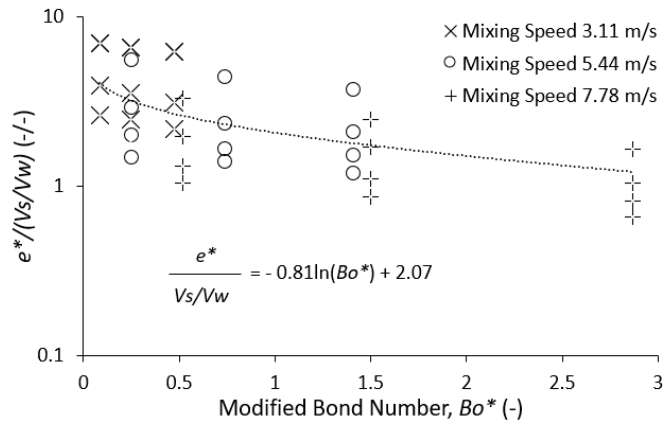

Figure 7. Empirical correlation between air entrapment normalized by initial solid volumetric concentration and modified Bond Number.

**Density Changes of mudflow mixtures containing air, water, and solid phases**

Besides air entrapment as an essential indicator of the mixing behavior of hydrophobic particles, water, and air, density change before and after the mixing process is necessary for mudflow models. The density reduction indicates how much density airless water and hydrophilic sand slurry will reduce due to the additional entrapment of air phases in the final hydrophobized sand surfaces.

Fig. 8a shows the effect of mixing time on density changes of the final mixture. The density reduction will reach a stable condition for the extended mixing time, which is shorter for coarser particles. After ruling out the time effect, mixing speed consistently impacts all types of sand particles (Fig. 8b). The average final density reduction normalized by the initial solid concentration is 17% for fine sand particles, 13% for medium particles, and 10% for coarse sand particles when the flow speed is fast at 7.78 m/s. We defined two bounding equations as a speed function to provide a range of estimation of density change normalized by initial solid concentration.

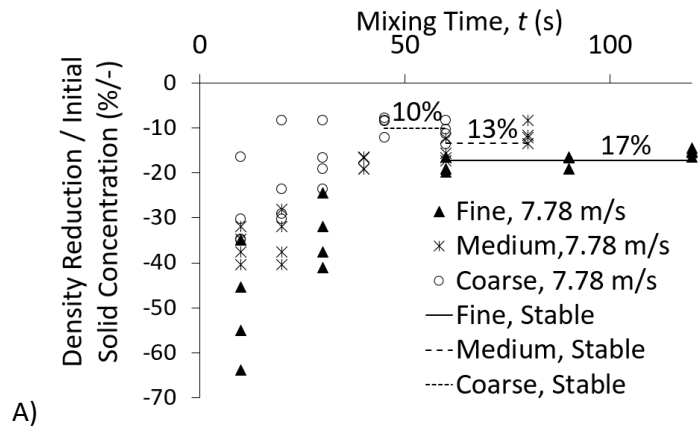

A)

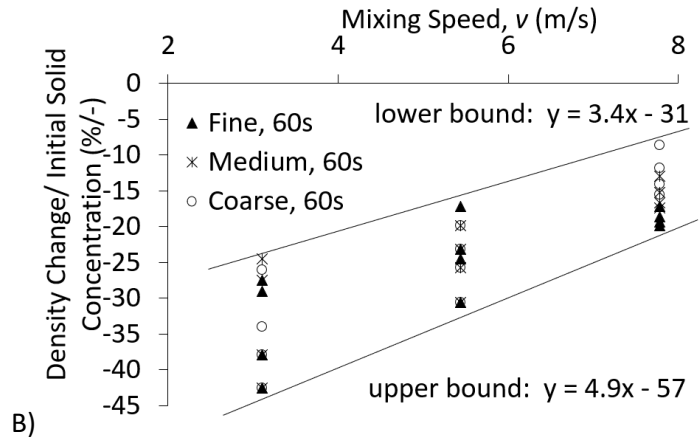

Figure 8. Density reduction because of A) mixing time and B) mixing speed.

Gravity drags the air-bubble-agglomerate downwards and, at the same time, buoyancy upwards. The final mixture arrangement separates into three parts, bottom-to-top: the mixed-phase settled layer, the pure water layer, and the top surface with a few floating agglomerates. As a result, a segregated mixture forms when mixing stops, as shown in detail for solid-dominated bottom layers in Fig. 9. Hydrophobic sand particles (shown in the box with a dotted line pattern) and small agglomerates (shown in the box with a dashed line pattern) settle in a compacted way at the bottom of the container. The largest agglomerates (shown in the box with a solid line pattern) are at the top. Only a few super-large agglomerates hover at the top of the free water layer.

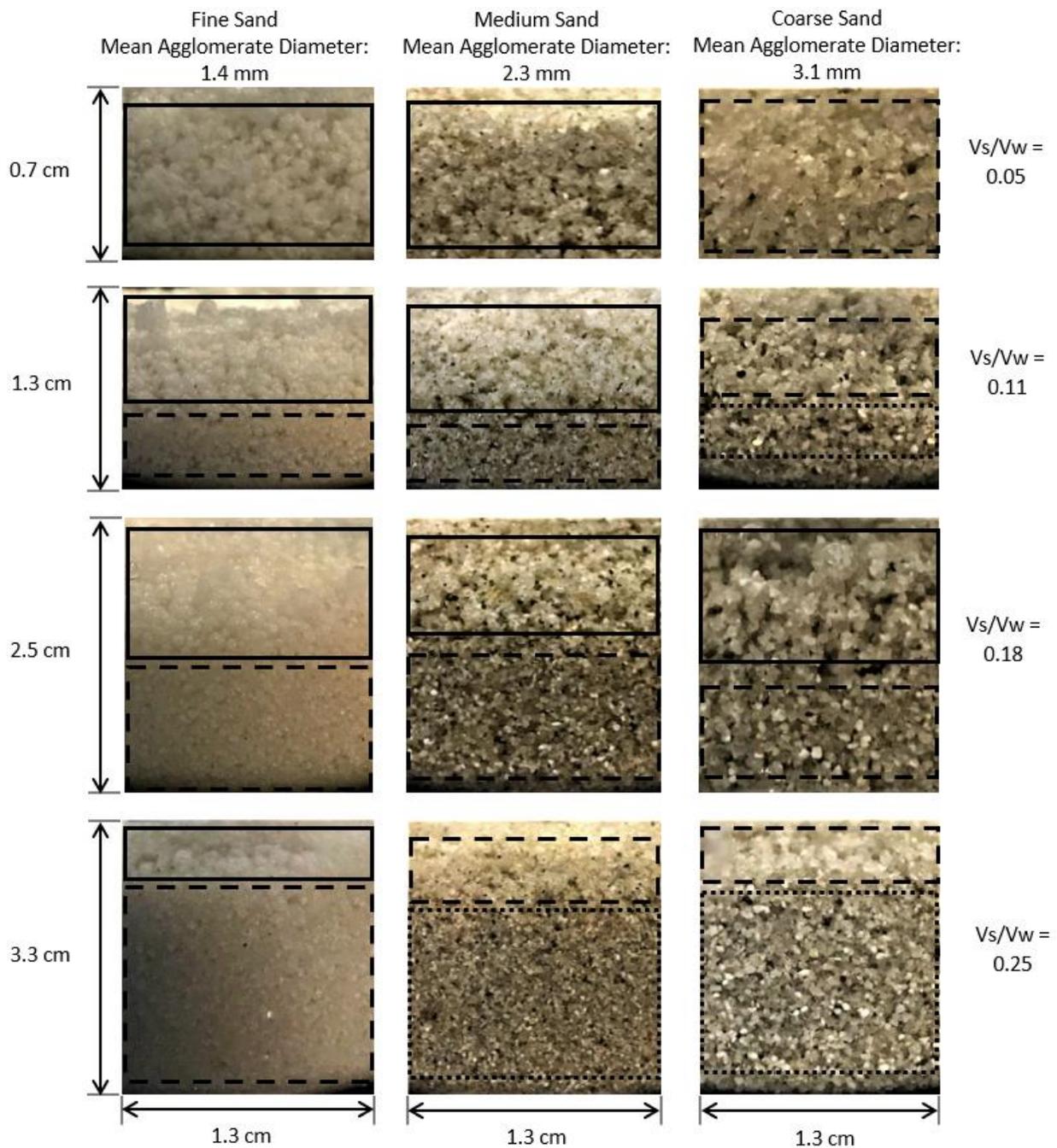

Figure 9. Layers of settlement arrangement for different phases.

**Discussion and Conclusions**

Mixing water, hydrophobic uniform sand, and air produces heterogeneous mixtures that entrap different amounts of air. Therefore, air trapping changes the mudflow mixture density and is crucial for defining overall flow and transport at larger scales. For the first time, this research investigates conditions that control air trapping within constraints of post-wildfire mudflow conditions. Fine, medium and coarse hydrophobic sands represent three different grains of sand as categorized from a geotechnical perspective. The air-trapping occurs due to the hydrophobic sand particles attaching to the air bubbles attracted into the mixture. As a result, air bubbles tend to flow upwards, while sand particles tend to settle in the opposite direction due to the Earth's gravity. Besides confirming previous findings that regardless of the availability of air phase in a closed system, the initial solid concentration has the strongest correlation with the final amount of entrapped air, we quantified and compared the effects of other coupled parameters, such as are mixing speed that is related to the mudflow downhill motion, mixing time, sand type, and variations in initial air-water-solids volumetric ratios. Furthermore, because of comprehensive experimental testing and analysis, the study highlights the effects and importance of sand type, differentiating mudflow mixture final composition in fine, medium, and coarse sands. We propose for the first time a forecasting formula for mudflow density under various conditions that include the air-trapping mechanism.

Specific new findings show that a longer mixing time gradually decreases the amount of entrapped air. In addition, mixing time is coupled with the average particle size, and coarser sand needs consistently less mixing time than fine sands to reach a steady volume of trapped air in the mixture at all investigated mixing speeds. Next, considering the mixing rate, air trapping into the mixture decreases as the mixing velocity increases and sand coarseness increases. Observing air bubbles and air-sand agglomerates can explain the variation and decrease of entrapped air in the mixture under faster, longer mixing and with coarser hydrophobic sands. Coarser sand forms larger agglomerates than finer sand. Agglomerates created from coarse sand have a more significant modified Bond Number, which means higher initial bubble shape irregularities. Agglomerate breakage is more prominent at a higher speed with a longer mixing time in coarser sand than in other sand. Furthermore, larger agglomerates deform more and subsequently break. Re, Ca, and We numbers lead to higher agglomerate susceptibility to breakage. We speculate that local turbulence and flow instabilities make particle-bubble interaction more unpredictable and increase the vulnerability of formed particle-bubble agglomerates towards breakage. Finally, although mixing blade geometry does not significantly affect the amount of entrapped air, we used three different blades to investigate its effect. Therefore, the results combine various blade geometry experiments in the same graphs.

Under given assumptions, the analysis provides a relationship between density change and parameters that can be back-calculated from field analysis, such as flow velocity and initial solid concentration.

**Methods**
**Energy and Mixing Speed**
A comprehensive mixing program is performed in controlled laboratory conditions to investigate the extent and forms of entrapped air after mixing with different speeds and spindles in gravity. The mixing process in a cup mimics the downhill flow and transport of the mixture rather than a linear setup. Therefore, the mixing speed translates to possible downhill mixture velocity using the work-energy principle in Eq. 3:

$$KE_{linear} = \frac{1}{2}mv^2 \qquad (3)$$

where *m* is the mixture mass, and *v* is the linear velocity of the mixture during a mudflow event. On the other side, the rotational kinetic energy in the experiments can be defined using the work-energy principle in Eq. 4:

$$KE_{rotational} = \frac{1}{2}I\omega^2 \qquad (4)$$

where ω is the angular velocity, *I* is the moment of inertia of the mixture. Fig. 10 compares the linear kinetic energy associated with downhill movement and the rotational kinetic energy related to mixing in a closed container. The rotational kinetic energy in laboratory mixing conditions shall equal linear kinetic energy in field conditions. For example, post-fire mudflow can move up to 30 kilometers per hour, or 8.3 meters per second (Cui et al., 2018).

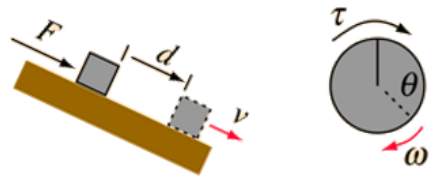

Figure 10. Relations between downhill velocity and experimental rotational agitation

**Experimental Setup**
Experiments use three types of sand: AFS 50/70 testing Ottawa silica fine sand (Fig. 11a, mean particle diameter: 0.25 mm), sieved 30/50 silica medium sand (Fig. 11b, mean particle diameter: 0.42 mm), and sieved 16/50 coarse sand (Fig. 11c, mean particle diameter: 0.59 mm). Sand and mixing blades are put in a 290 ml volume commercial cup with an overall height and diameter of 75 and 70 mm.

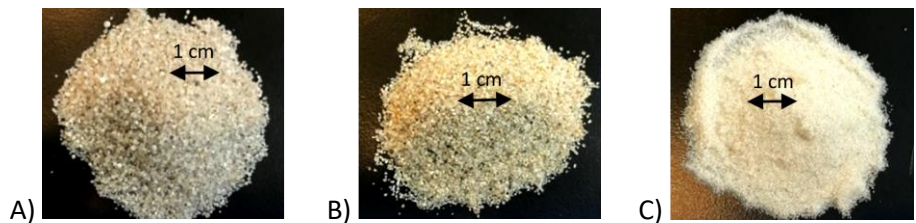

Figure 11. Examples of A) coarse sand, B) medium sand, and C) fine sand are used in experiments.

Figs. 12a-c show the pictures of the physical appearance of mixing blades used in the experiments. Type I and Type III blades have four leaves, while Type II has three leaves. Table 1 shows a detailed geometric comparison among three types of mixing blades. Our investigation demonstrates that blade geometry does not affect air entrapment (Fig. 13). However, when combining blade geometry with initial solid volumetric concentration or type of sand particles, the latter factors strongly dominate air entrapment behavior.

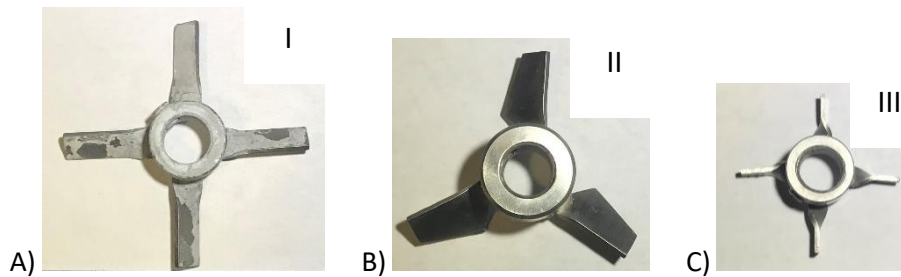

Figure 12 Mixing blade geometries considered for the experiments: A) Type I with 5 cm span, B) Type II with 4.5 cm span, and C) Type III with 4 cm span.

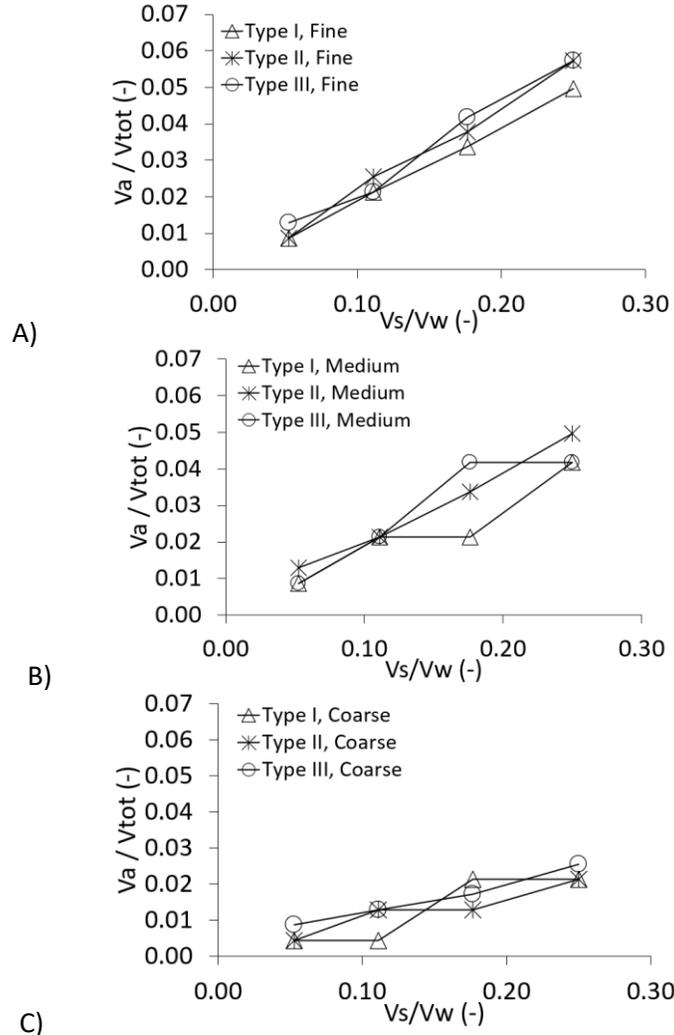

Figure 13. Effect of mixer geometry on the ability of air entrapment.

We apply hydrophobic coating on the sand by following three steps. First, we wash untreated sand underwater until fine dust is removed. Next, we put the washed sand in the oven, heating for at least 24 hours and 100 °C until thoroughly dried. Second, after the sand is cooled down, we merge sand in a mixed solution of Triethoxy-n-octylsilane ($C_{14}H_{32}O_3Si$) and isopropyl alcohol ($CH_3CHOHCH_3$) for at least 48 hours. The solution contains 10% triethoxy-n-octylsilane and 90% isopropyl alcohol by volume. Last, we take the sand out of the chemicals and cleaned them underwater. We put the cleaned sand back in the oven again, heating for at least 24 hours until the sand was dry. Table 2 summarizes the contact angles for each type of sand after hydrophobicity treatment.

    We also vary fluid mixing speed, mixing time, and initial volumetric concentration of sand to the water. We apply three different mixing speeds for all types of sand particles at 3.11 m/s, 5.44 m/s, and 7.78 m/s. Mixing time varies from 10 seconds to 120 seconds for different sand particles. The initial volumetric ratio of sand to water is determined by measuring the weight (mass) and sand's specific gravity ($G_S$). Specific gravity is the ratio of the weight of the given volume of aggregates to the weight of an equal volume of water. $G_S$ for silica sand is usually 2.65, which does not vary too much. Equation (5) shows the relationship between the mass of sand ($M_S$) and the volume of sand solids ($V_S$):

$$M_S = \frac{G_S \cdot \gamma_W \cdot V_S}{g} \qquad (5)$$

where, $\gamma_w$ is the unit weight of water (9.8 kN/m³), $g$ is the gravitational acceleration (9.8 m/s²). Finally, the same method is applied to obtain other volumetric ratios/concentrations.

The custom mixing system uses a 1HP motor attached to the mixer blade. Fig. 14a shows a picture of the mixing station. The motor is fixed on the top of the 38 by 64 cm frame. At the other end, the motor shaft connects to the mixer blade via several different shafts and bearings to fit the mixer blade's borehole size. We put the mixing container at the very bottom of the system. A variable-frequency drive (VFD) controls the speed of the motor. Figs. 14b and 14c show the top and front views of the mixing container and mixing blade dimensions, respectively.

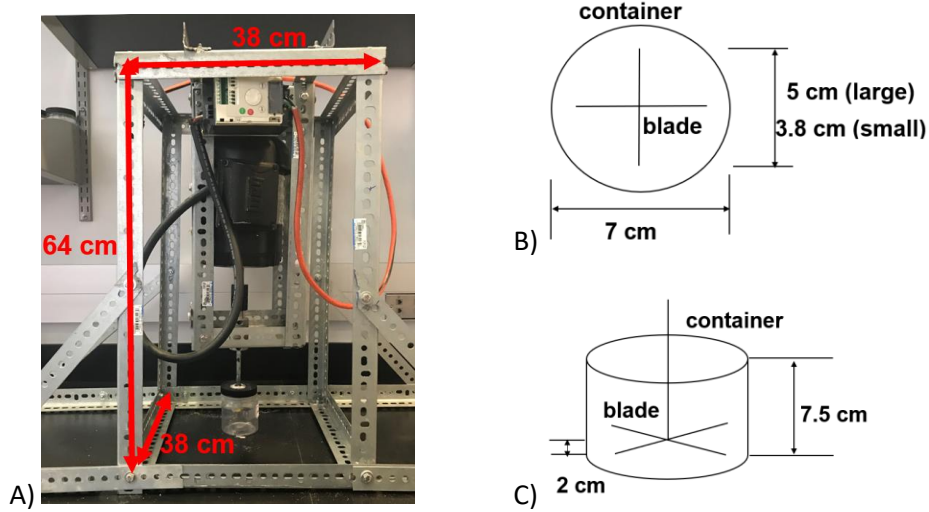

Figure 14. Illustrations for: A) mixing station; B) plan view and dimension of the mixing container; C) front view and dimension of mixing container

To prevent water or air leakage, we apply a sealant to all possible gaps around the container. Before mixing, a certain amount of sand was at the bottom of the container, followed by pouring water above the sand layer. Air exists naturally within the hydrophobic sand layer and above the water layer within the container. Air bubbles generate once the mixer blade starts rotating.

**Dimensionless Quantities**
The Bond Number (*Bo*) is the ratio of inertia to capillary force used to assess the stability of a bubble or droplet within another type of surrounding fluid. A lower *Bo* indicates that surface tension dominates the general behavior of a bubble or droplet. *Bo* describes a two-phase system; however, Schulze (1982) proposed a modified Bond Number (*Bo**). *Bo** is the ratio of all detaching forces to attaching forces of a particle-bubble agglomerate within some surrounding fluid. If Bo* is less than one, particles will likely stay connected to the bubble, and if Bo* is larger than one, particles will probably detach from the bubble:

$$B_o^* = \frac{F_g - F_b + F_a + F_p}{F_c} \qquad (6)$$

where, $F_g$ is the gravity, $F_b$ is the buoyancy, $F_a$ is the centrifugal force, $F_p$ is the net pressure force, and $F_c$ is the capillary force. After replacing the expressions for the forces, $Bo^*$ can be simplified:

$$B_o^* = \frac{d_p^2(g\Delta\rho + \rho_p b_m) - d_p \sigma \cos^2\left(\frac{\theta}{2}\right)}{6\sigma \sin^2\left(\frac{\theta}{2}\right)} \quad (7)$$

$$b_m = \frac{\bar{u}^2}{r} \quad (8)$$

where $d_p$ is the particle diameter (m), $g$ is the gravity acceleration (m/s$^2$), $\Delta\rho$ is the density difference between particle and liquid (kg/m$^3$), $\rho_p$ is the particle density (kg/m$^3$), $b_m$ is the eddy turbulent acceleration (m/s$^2$), σ is the surface tension (N/m), θ is the contact angle (degree), $\bar{u}$ is the mean fluctuating velocity (i.e., assumed to be mixing speed) (m/s), and r is the rotating length scale (m). Table 3 summarizes the parameters required to calculate $Bo^*$.

**Acknowledgment**
The U.S. National Science Foundation supported this work, Division of Chemical, Bioengineering, Environmental, and Transport Systems, Directorate for Engineering (grant number NSF/CBET 2025643). In addition, we acknowledge Mahta Movasat for giving constructive feedback on the paper.

**Author information**
**Affiliations**
Department of Structural Engineering, University of California, San Diego, USA
Wenpei Ma (ORCID: https://orcid.org/0000-0002-6397-529X) & Ingrid Tomac (e-mail: itomac@ucsd.edu, https://orcid.org/0000-0002-7969-9525)
**Contributions**
W.M wrote the main manuscript and prepared figures and tables. I.T provided reviews and corrections.
**Corresponding author**
Correspondence to Wenpei Ma (w6ma@eng.ucsd.edu).

**Ethics declarations**
**Competing interests**
The authors declare no competing interests.

**References**
1. Brabcová, Z., Karapantsios, T., Kostoglou, M., Basařová, P., Matis,K. Bubble–particle collision interaction in flotation systems. Colloids and Surfaces A: Physicochemical and Engineering Aspects **473**, 95-103 (2015).
2. Bull, W. B. Alluvial-fan deposits in western Fresno County, California. The Journal of Geology **71**, 243–251 (1963).
3. Cannon, S.H, Krikham, R.M. & Parise, M. Wildfire-related debris-flow initiation processes, Strom King Mountain, Colorado. Geomorphology 39, 171-188 (2001).
4. Cannon, S.H. et al. Storm rainfall conditions for floods and debris flows from recently burned areas in southwestern Colorado and southern California. Geomorphology **96**, 250-269 (2008).
5. Cervantes-Álvarez, A.M. et al. Air entrainment and granular bubbles generated by a jet of grains entering water. Journal of Colloid and Interface Science 574, 285-292 (2020).

6. Conedera, M. et al. Consequences of forest fires on the hydrogeological response of mountain catchments: a case study of the Riale Buffaga, Ticino, Switzerland. Earth Surface Processes and Landforms, 28, 117-129 (2003).
7. Crandell, D. R. &Waldron, H.H. A Recent volcanic mudflow of exceptional dimensions from Mount Rainier, Washington. American Journal of Science 254, 349-362 (1956).
8. Cui, Y., Cheng, D. & Chan, D. Investigation of Post-Fire Debris Flows in Montecito. ISPRS International Journal of Geo-Information **8**, 5 (2018).
9. DeBano, L. F., Rice, R. M., Eugene, C. C. Soil heating in chaparral fires: effects on soil properties, plant nutrients, erosion, and runoff. Res. Paper PSW-RP-145. Berkeley, CA: U.S. Department of Agriculture, Forest Service, Pacific Southwest Forest and Range Experiment Station. 21. (1979).
10. DeBano, L.F.. Water repellent soils: a state-of-the-art. USDA Forest Service General Technical Report PS W-46, 21 (1981).
11. DeBano, L.F.. The effect of fire on soil properties. Proceedings management and productivity of western-Montane. Forest Soils. (1991).
12. DeBano, L.F.. The role of fire and soil heating on water repellency in wildland environments: a review. Journal of Hydrology, **231–232**, 195-206 (2000).
13. Dunkerley D. Rainfall intensity in geomorphology: Challenges and opportunities. Progress in Physical Geography: Earth and Environment 45, 488-513 (2021).
14. Eskanlou, A., Chegeni, M. H., Khalesi, M. R., Abdollahy, M., Huang, Q. Modeling the bubble loading based on force balance on the particles attached to the bubble. Colloids and Surfaces A: Physicochemical and Engineering Aspects, 582, 123892 (2019).
15. Fielden, M. L., Hayes, R. A., Ralston J. Surface and Capillary Forces Affecting Air-bubble–Particle Interactions in Aqueous Electrolyte. Langmuir,12 (15), 3721-3727 (1996).
16. Fornasiero, D. & Filippov, L. O. Innovations in the flotation of fine and coarse particles. Journal of Physics.: Conference Series **879**, 012002 (2017).
17. Gai, G., et al. Particles-induced turbulence: A critical review of physical concepts, numerical modelings and experimental investigations. *Theoretical and Applied Mechanics Letters* **10**, 241-248 (2020).
18. Gao, Y., Evans, G. M., Wanless, E. J., Moreno-Atanasio, R. DEM simulation of single bubble flotation: Implications for the hydrophobic force in particle–bubble interactions. Advanced Powder Technology, **25**, 1177-1184 (2014).
19. Garoosi, F. et al. Experimental and numerical investigations of dam break flow over dry and wet beds. International Journal of Mechanical Sciences 215, 106946 (2022).
20. Gillies, G., Kappl, M., Butt, H.-J. Direct measurements of particle–bubble interactions. Advances in Colloid and Interface Science **114–115**, 165-172 (2005).
21. Ishida, N. Direct measurement of hydrophobic particle–bubble interactions in aqueous solutions by atomic force microscopy: Effect of particle hydrophobicity. Colloids and Surfaces A: Physicochemical and Engineering Aspects **300**, 293-299 (2007).
22. Johnson, D. J., Miles, N. J., Hilal, N. Quantification of particle–bubble interactions using atomic force microscopy: A review. Advances in Colloid and Interface Science, **127**, 67-81 (2006).
23. Kean, J.W., Staley, D.M., Cannon, S.H. In situ measurement of post-fire debris flows in southern California: Comparisons of the timing and magnitude of 24 debris-flow events with rainfall and soil moisture conditions. Journal of Geophysical Research Earth Surface, 116, F04019 (2011).
24. Lee, S. H.-H. & Widjaja, B. Phase concept for mudflow based on the influence of viscosity. *Soils and Foundations* **53**, 77-90 (2013).
25. Liu, T.y., Schwarz, M.P. CFD-based modelling of bubble-particle collision efficiency with mobile bubble surface in a turbulent environment. International Journal of Mineral Processing, **90**, 45-55 (2009).

26. Lu, S. Hydrophobic interaction in flocculation and flotation 3. Role of hydrophobic interaction in particle—bubble attachment. Colloids and Surfaces **57**, 73-81 (1991).
27. Maxwell, R., Ata, S., Wanless, E.J., Moreno-Atanasio, R. Computer simulations of particle–bubble interactions and particle sliding using Discrete Element Method. Journal of Colloid and Interface Science **381**, 1-10 (2012).
28. Neary, D. G., Ryan, K. C., DeBano, L. F.. Wildland fire in ecosystems: effects of fire on soils and water. Gen. Tech. Rep. RMRS-GTR-42-vol.4. Ogden, UT: U.S. Department of Agriculture, Forest Service, Rocky Mountain Research Station **250**, (2005).
29. Nguyen, A. V., Schulze, H. J., Ralston, J. Elementary steps in particle—bubble attachment. International Journal of Mineral Processing **51**, 183-195 (1997).
30. Ong, X.Y. et al. On the formation of dry granular jets at a liquid surface. Chemical Engineering Science 245, 116958 (2021).
31. Phan, C. M., Nguyen, A. V., Miller, J. D., Evans, G. M., Jameson, G. J. Investigations of bubble–particle interactions. International Journal of Mineral Processing **72**, 239-254 (2003).
32. Preuss, M. & Butt, H.-J. Direct measurement of particle-bubble interactions in aqueous electrolyte: Dependence on surfactant. Langmuir **14**, 3164-3174 (1998).
33. Pyke, B., Fornasiero, D., Ralston, J. Bubble particle heterocoagulation under turbulent conditions. Journal of Colloid and Interface Science **265**, 141-151 (2003).
34. Römkens, M.J.M., Prasad, S.N. & Gerits, J.J.P. Soil erosion modes of sealing soils: a phenomenological study. Soil Technology 11, 31-41 (1997).
35. Schulze, H.J. Dimensionless Number and Approximate Calculation of the Upper Particle Size of Floatability in Flotation Machines. International Journal of Mineral Processing 9, 321-328 (1982).
36. Schulze, H.J. Hydrodynamics of Bubble-Mineral Particle Collisions. Mineral Processing and Extractive Metallurgy Review **5**, 43-76 (1989).
37. Sheng, L.T et al. A two-phase model for dry density-varying granular flows. Advanced Powder Technology 24, 132-142 (2013).
38. Song, D. et al. Impact dynamics of debris flow against rigid obstacle in laboratory experiments. Engineering Geology 291, 106211 (2021).
39. Suhr, J.L., Jarrett, A.R. & Hoover, J.R. The effect of soil air entrapment on erosion. Transactions of the ASAE 27, 0093-0098 (1984).
40. Tanaka, Y. et al. Bench-scale experiments on effects of pipe flow and entrapped air in soil layer on hillslope landslides. Geosciences 9, 138 (2019).
41. Verrelli, D. I., Koh, P. T. L., Nguyen, A. V. Particle–bubble interaction and attachment in flotation. Chemical Engineering Science **66**, 5910-5921 (2011).
42. Wang, G., Evans, G. M., Jameson, G. J. Bubble–particle detachment in a turbulent vortex I: Experimental. Minerals Engineering **92**, 196-207 (2016).

Table 1 Detailed geometrical description of each type of mixing blade

|  | Type I | Type II | Type III |
|---|---|---|---|
| Commercial Name | Mixer Direct's 2" Lab Hydrofoil Blade | Mixer Direct's 2" Lab Axial Flow Turbine Blade | Mixer Direct's 1.5" Axial Flow Turbine Blade |
| Blade Diameter (inch) | 2 | 2 | 1.5 |
| Blade Diameter (cm) | 5.1 | 5.1 | 3.8 |
| Number of Leaves | 4 | 3 | 4 |
| Leaf Separation Angle (°) | 90 | 120 | 90 |
| Leaf Inclination Angle (°) | 45 | 45 | 90 |
| Leaf Length (cm) | 1.9 | 1.9 | 1.3 |
| Leaf Width (cm) | 0.5, uniform | 1 to 0.5, reducing | 0.5, uniform |
| Leaf Thickness (cm) | 0.16 | 0.16 | 0.16 |

Table 2 Contact angles for different types of hydrophobic sand

| Sand Type | Contact Angle (°) |
|---|---|
| Hydrophobic Fine Sand | 116 |
| Hydrophobic Medium Sand | 100 |
| Hydrophobic Coarse Sand | 95 |

Table 3 Parameters for calculating modified Bond number of agglomerates

|  |  | FINE | | | MEDIUM | | | COARSE | | |
|---|---|---|---|---|---|---|---|---|---|---|
| Particle Diameter | (mm) | 0.25 | 0.25 | 0.25 | 0.4 | 0.4 | 0.4 | 0.59 | 0.59 | 0.59 |
| Gravity Acceleration | (m/s$^2$) | 9.80 | 9.80 | 9.80 | 9.80 | 9.80 | 9.80 | 9.80 | 9.80 | 9.80 |
| Particle and Liquid Density Difference | (kg/m$^3$) | 1600 | 1600 | 1600 | 1600 | 1600 | 1600 | 1600 | 1600 | 1600 |
| Particle Density | (kg/m$^3$) | 2600 | 2600 | 2600 | 2600 | 2600 | 2600 | 2600 | 2600 | 2600 |
| Mean Fluctuation Velocity | (m/s) | 3.11 | 5.44 | 7.78 | 3.11 | 5.44 | 7.78 | 3.11 | 5.44 | 7.78 |
| Rotating Length Scale | (m) | 0.07 | 0.07 | 0.07 | 0.07 | 0.07 | 0.07 | 0.07 | 0.07 | 0.07 |
| Eddy Turbulent Acceleration | (m/s$^2$) | 138 | 423 | 865 | 138 | 423 | 865 | 138 | 423 | 865 |
| Surface Tension | (N/m) | 0.06 | 0.06 | 0.06 | 0.07 | 0.07 | 0.07 | 0.07 | 0.07 | 0.07 |
| Contact Angle | (°) | 116 | 116 | 116 | 100 | 100 | 100 | 95 | 95 | 95 |
| Bo* | (-) | 0.09 | 0.25 | 0.52 | 0.25 | 0.74 | 1.50 | 0.47 | 1.41 | 2.87 |